\documentclass[submission]{eptcs}
\usepackage{graphicx}
\usepackage{afterpage}
\providecommand{\event}{FTSCS 2012} 

\title{Model Checking with Program Slicing Based on Variable Dependence Graphs}
\author{
Masahiro Matsubara \qquad Kohei Sakurai \qquad Fumio Narisawa
\institute{Hitachi Research Laboratory, Hitachi, Ltd.\\Japan}
\email{[masahiro.matsubara.td|kohei.sakurai.cp|fumio.narisawa.ks]@hitachi.com}
\and
Masushi Enshoiwa \qquad Yoshio Yamane
\institute{Hitachi Advanced Digital, Inc.\\Japan}
\and
Hisamitsu Yamanaka
\institute{Hitachi Automotive Systems, Ltd.\\Japan}
}
\def\titlerunning{Model Checking with Program Slicing Based on Variable Dependence Graphs}
\def\authorrunning{M.Matsubara, K.Sakurai, F.Narisawa, M.Enshoiwa, Y.Yamane \& H.Yamanaka}
\begin{document}
\maketitle

\begin{abstract}
In embedded control systems, the potential risks of software defects have been increasing because of software complexity which leads to, for example, timing related problems.
These defects are rarely found by tests or simulations.
To detect such defects, we propose a modeling method which can generate software models for model checking with a program slicing technique based on a variable dependence graph.
We have applied the proposed method to one case in automotive control software and demonstrated the effectiveness of the method. Furthermore, we developed a software tool to automate model generation and achieved a 35\%
decrease in total verification time on model checking.
\end{abstract}

\section{Introduction}

In embedded control systems, potential risks to system safety have been increasing because programs are getting larger due to electrification, enhancement of diagnosis, etc. Functional safety standards such as IEC 61508 or ISO 26262 have been established to ensure these systems do not fall into dangerous situations. However, the problem of test coverage still remains. Furthermore, there are corner cases which are difficult to be found with usual tests or simulations.

For example, hardware malfunctions could cause software faults, so it is not sufficient to test the software only. Inspection techniques to test combinations of hardware and software such as HILS (Hardware in-the-Loop Simulation) have been in practical use, but failures caused by hardware malfunctions or timing problems caused by software interruptions are difficult to be detected because there could be large number of test cases.

To solve these problems, model checking is applied. In model checking, all of the state transitions of the system are fully searched to detect corner cases including timing problems. Implementation bugs can also be found if models are made out of the source code. However, the problem in applying model checking is well known: a state explosion that verification does not complete. To avoid a state explosion, the scope of verification has to be limited. However, too limited a model misses some causes of malfunctions, so that the part of the source code to be verified has to be determined appropriately.

We are trying to adopt model checking to the development of automotive control software, using it to find software defects which occur rarely on real systems by means of modeling both the software from the source code and the hardware. To solve the state explosion problem mentioned above, we introduced a program slicing technique \cite{Weiser:1981} based on a variable dependence graph which enables an flexible adjustment of the boundaries between the source code to be modeled as software model in detail and the external environment to the software model. Also we introduced external environment models which include unsteady behavior or faults in them to detect malfunctions caused by interactions of hardware and software.

The contributions of this paper are as follows:
\begin{itemize}
\item A practical modeling method to verify a large scale embedded control program. The main point of the method is the slicing technique based on a variable dependence graph.
\item The algorithm for analysis of variable dependence graph, which could be adapted to the analysis of system dependence graph.
\item An experimental evaluation of the tool which automates the above method.
\end{itemize}

The rest of this paper is structured as follows: Section \ref{sec:modeling} presents the modeling method and the slicing technique. Section \ref{sec:study} presents an example case where the method is applied, to show the usefulness of the method. Section \ref{sec:tool} presents the tool which automates the method. Section \ref{sec:related} presents the related work. Finally, Section \ref{sec:conclusion} concludes this paper by summarizing its main points.

\section{Modeling Method}
\label{sec:modeling}

\subsection{System Modeling}
\begin{figure}[tb]
 \begin{center}
  \includegraphics[viewport=30 30 440 750, scale=0.55, angle=270, clip]{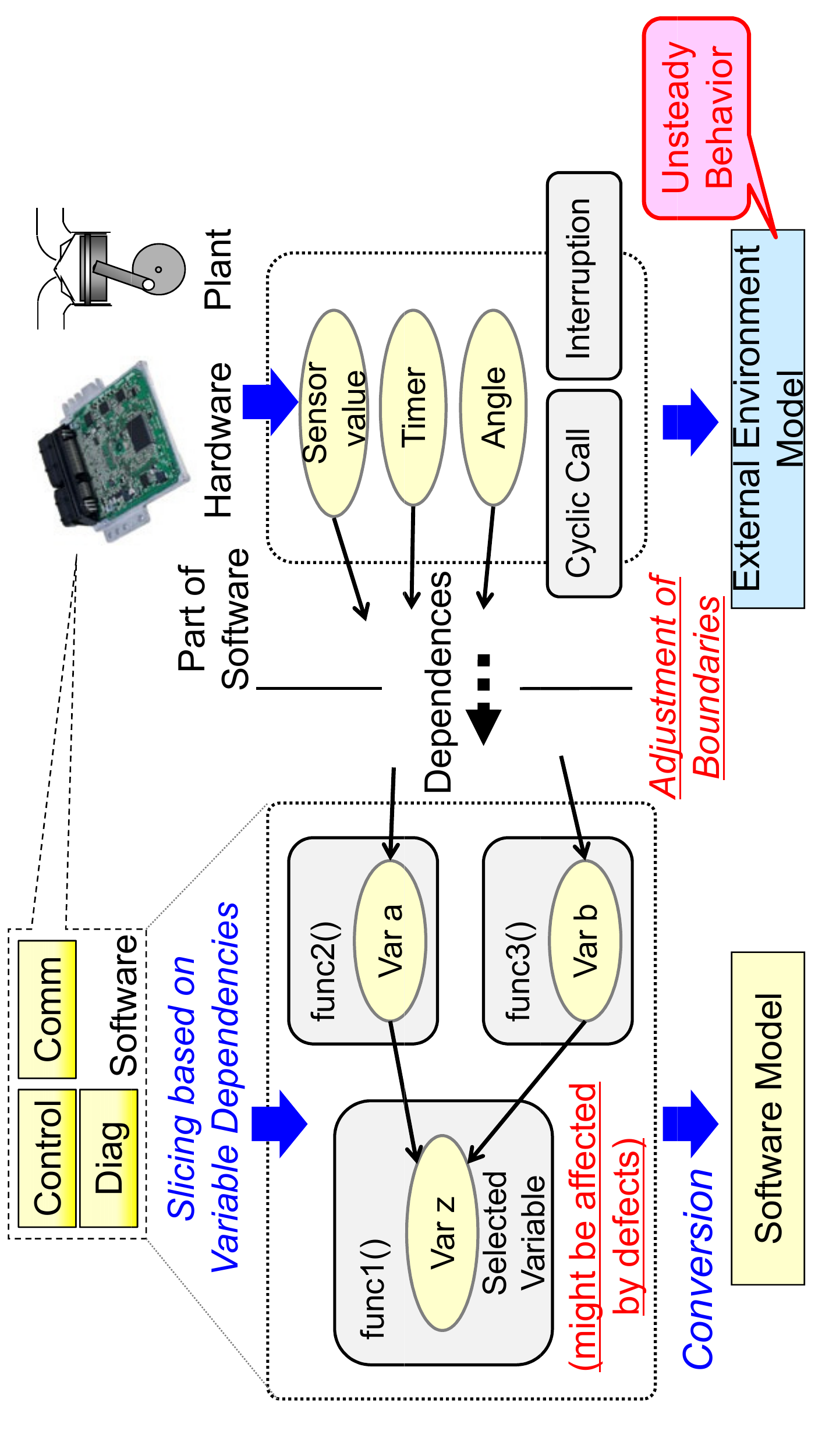}
  \caption{Modeling Method for Embedded Control System}\label{fig:method}
 \end{center}
\end{figure}

The way of modeling an embedded control system is shown in Figure \ref{fig:method}. To find defects in implementations, the software model of the target system is converted from the source code which is related to the selected variables by a verifier. Those selections are done in accordance with a verification point. The code to be converted is sliced out from the whole program to avoid a state explosion, using a slicing technique based on a variable dependence graph (see \ref{sec:technique}).

On the other hand, the hardware, controlled instruments (if needed for verification), and a part of the software are modeled as external environments. These external environment models can have unsteady behavior or faults if they are concerned with the content to be verified.

The slicing technique based on the variable dependence graph makes it easy to adjust the boundaries between the software models and the external environment models. The reason that such adjustments are needed is that the appropriate size of the software model differs depending on the verification points. Too large a model leads to a state explosion, and too small a model can miss the cause of a malfunction.
 
\begin{figure}[tb]
 \begin{center}
  \includegraphics[viewport=20 30 500 315, scale=0.6, angle=0, clip]{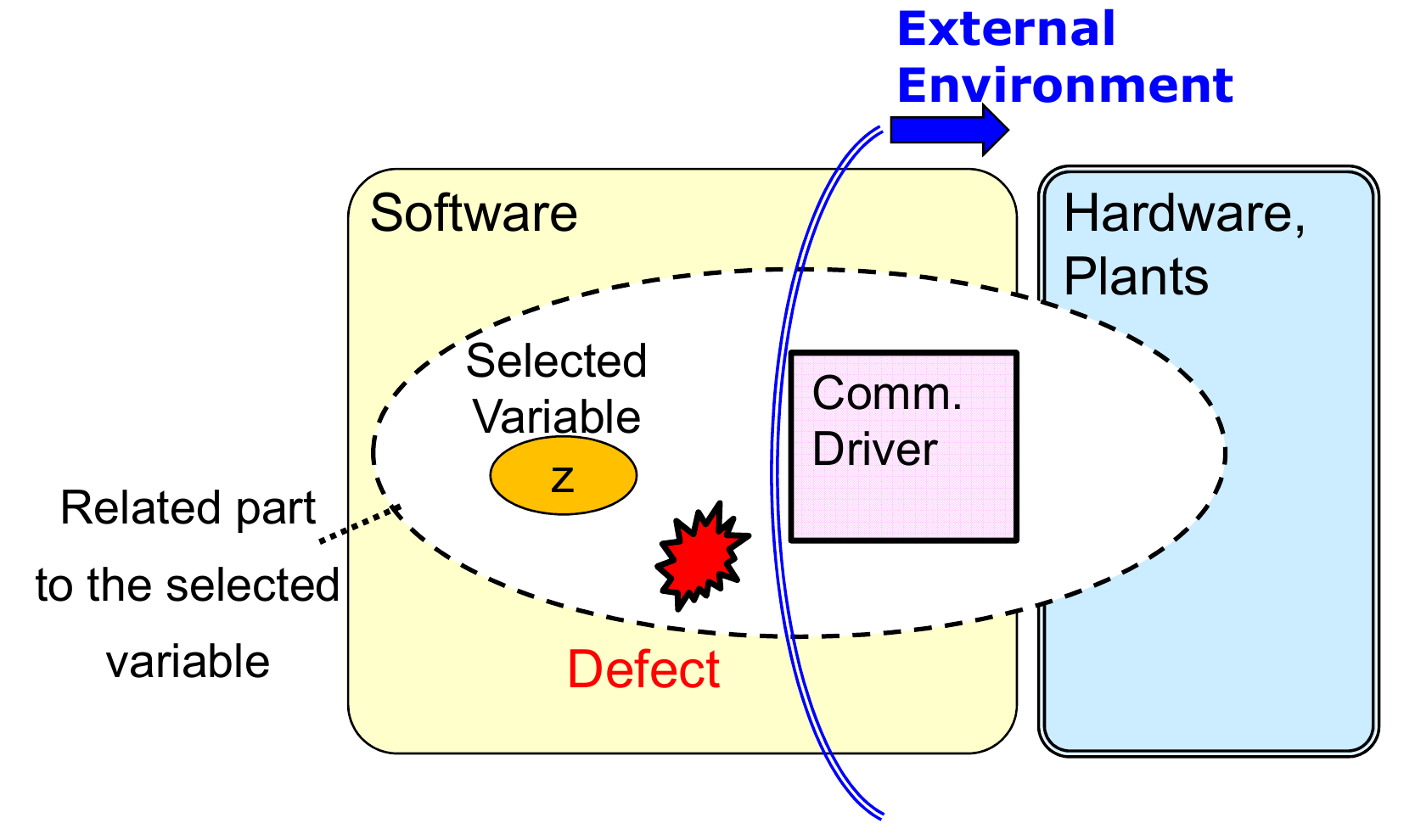}
  \caption{Abstraction of Boundary between Software and Hardware}\label{fig:abstraction}
 \end{center}
\end{figure}

An example of a boundary adjustment is shown in Figure \ref{fig:abstraction}. There is a communication driver between the software and the hardware, and this driver has complex processing. It is better to omit the driver to reduce the state number of the software model, if the driver has no relation to a malfunction. In this case, it is appropriate to treat the driver as a part of the external environment, and simplify its processing.
This method could make the modeling highly automated but not fully automated. The use of design knowledge is expected for modeling, so that the scope of the software model has to be changed according to the verification point.

\subsection{Slicing Technique}
\label{sec:technique}
 The data to be used for slicing is usually a program dependence graph (PDG) or a system dependence graph (SDG) \cite{Horwitz:1988}. The criteria for slicing include a start statement, an end statement, and variables in the end statement.

\begin{figure}[tb]
 \begin{center}
  \includegraphics[viewport=30 30 455 660, scale=0.7, angle=270, clip]{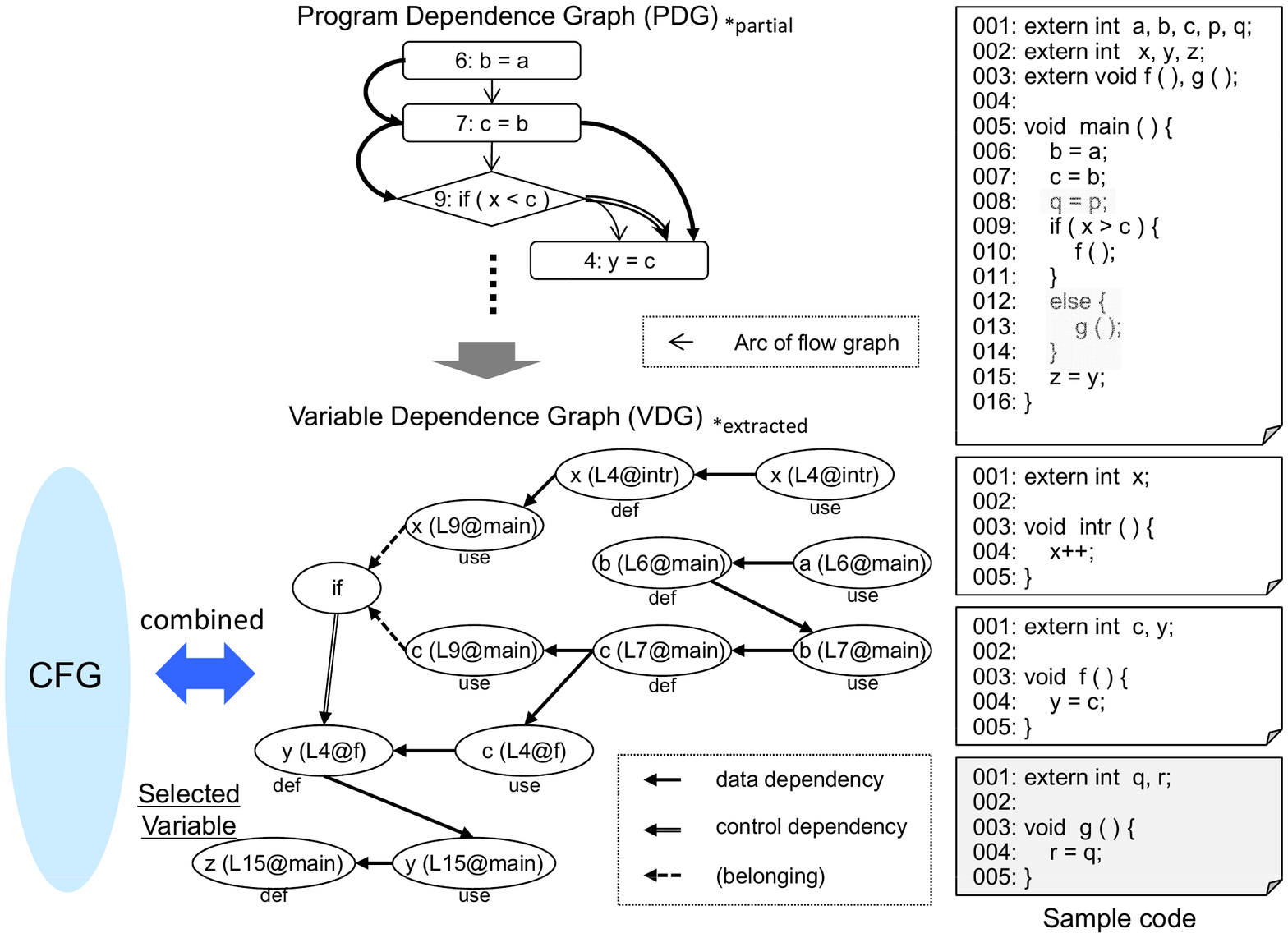}
  \caption{Variable Dependence Graph of Sample Code}\label{fig:VDG}
 \end{center}
\end{figure}

In this method, the variable dependence graph (VDG) in Figure \ref{fig:VDG} is used as data for slicing, which is a kind of data flow to connect dependence among the variables. VDG is equivalent to PDG (or SDG) except for the data unit (a statement in PDG), but it is more suitable for expressing dependencies in a tree format. Tree format is better to pursue related variables and code sequentially, and to adjust the boundaries between code converted to the software model and code treated as external environments.
In PDG, data dependence and control dependence are used as an edge. On the other hand, in VDG, dependence brought by an assignment which is similar to data dependence is also used as an edge. Nodes in VDG are variables distinct with positions and execution paths in the source code.

There are two directions in VDG trees. In one, the root node is the goal of the dependencies in the tree (we call it the goal tree): in the other, the root node is the start (the start tree). These root nodes are one side of the slicing criteria, and leaf nodes including those which are on a boundary adjusted by a verifier are another side.

\begin{figure}[tb]
 \begin{center}
  \includegraphics[viewport= 30 50 480 270, scale=0.8, angle=0, clip]{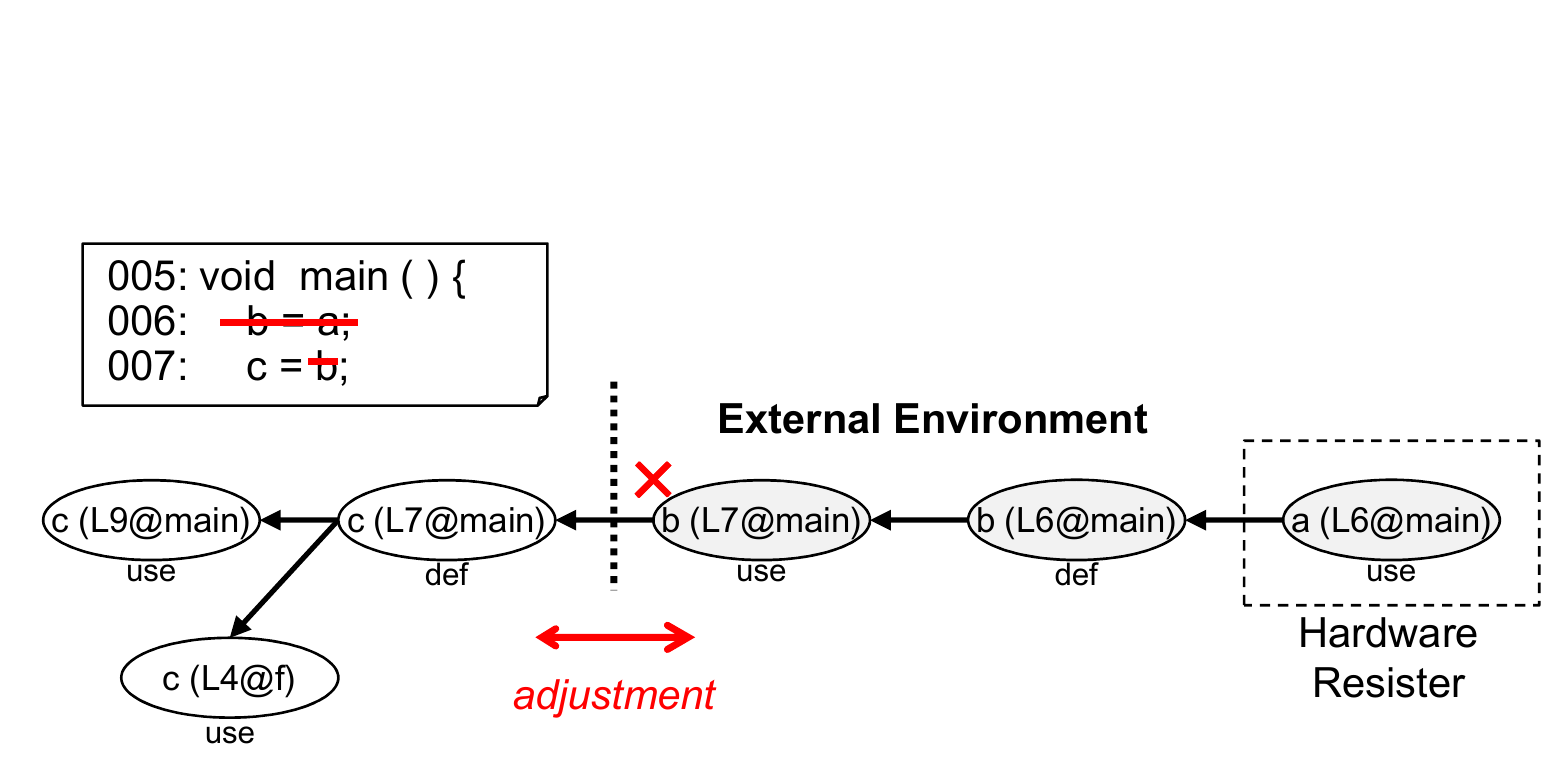}
  \caption{Adjustment Boundary on VDG}\label{fig:adjustment}
 \end{center}
\end{figure}

The steps of slicing are as follows:
\begin{enumerate}
\item Select variables which are related to the verification point in the source code of the target system.
 These variables are those which could be affected by defects, and are written in properties or assertions.
\item Analyze and extract the VDG goal / start trees from the source code, where the root nodes are the variables selected in step 1.
\item Adjust the boundaries in the VDG trees. (Figure \ref{fig:adjustment})
\item Execute slicing to extract code which has a relation to the variables in the VDG trees extracted in Step 2 and adjusted in Step 3.
\end{enumerate}

In step 3, a variable on a boundary becomes interface between the software model and a external environment model. It is possible to set a boundary between functions by setting the boundary between variables which belong to different functions.

In step 4, code to be extracted are not only statements which include a variable in VDG directly, but also declarations of the variables, functions and control statements which include or relates to the above statements in control flow.

Statements (except function calls) which do not include a variable cannot be extracted, so that these statements have to be picked up additionally on a control flow if needed.

The benefits of this slicing technique are as follows:
\begin{itemize}
\item High accuracy of slicing, equal to PDG-based slicing.
\item Easy to adjust boundaries between software models and external environments, especially in the tree expression of variable dependencies.
\item Interface variables between software models and external environments are clear. Information on the interface is important to make or select external environment models.
\item Variables whose range has to be changed are clear when input / output values have data mapping.
\end{itemize}

The sliced code is converted into software models in the language of the model checker, such as Promela of SPIN \cite{Holtzman:1997}. This conversion is done via AST (Abstract Syntax Tree) according to the predetermined rules.

\subsection{External Environment Models}
External environment models are necessary to detect malfunctions caused by interactions between the hardware and software. Specifically, they set every possible combination of input values to software models, or limit those combinations to reduce the state number according to the specification of the external environment. They also simulate the behavior outside of the software so that it becomes possible to judge whether a property for the external environment is held or not.

External environment models are written manually. When a verifier makes a new model or selects one from the existing models of the external environment, information about the interface variable given in VDG is useful. Usually, input values to a software model or determinants of them are set with randomness. Processes are added to external environment models.

Furthermore, a verifier can put unsteady behavior or faults into the external environment models which occur at random realized with non-determinism. If there is a malfunction when an external environment model has unsteady behavior, and vice versa, that behavior seems to be one of the factors which cause the malfunction, so that a verifier could analyze a counter example from that point of view.

\section{Case Study}
\label{sec:study}
In this section, we apply the modeling method described in section \ref{sec:modeling} to show its usefulness.

\subsection{Example Case}
\begin{figure}[tb]
 \begin{center}
  \includegraphics[viewport=30 30 485 285, scale=0.5, angle=0, clip]{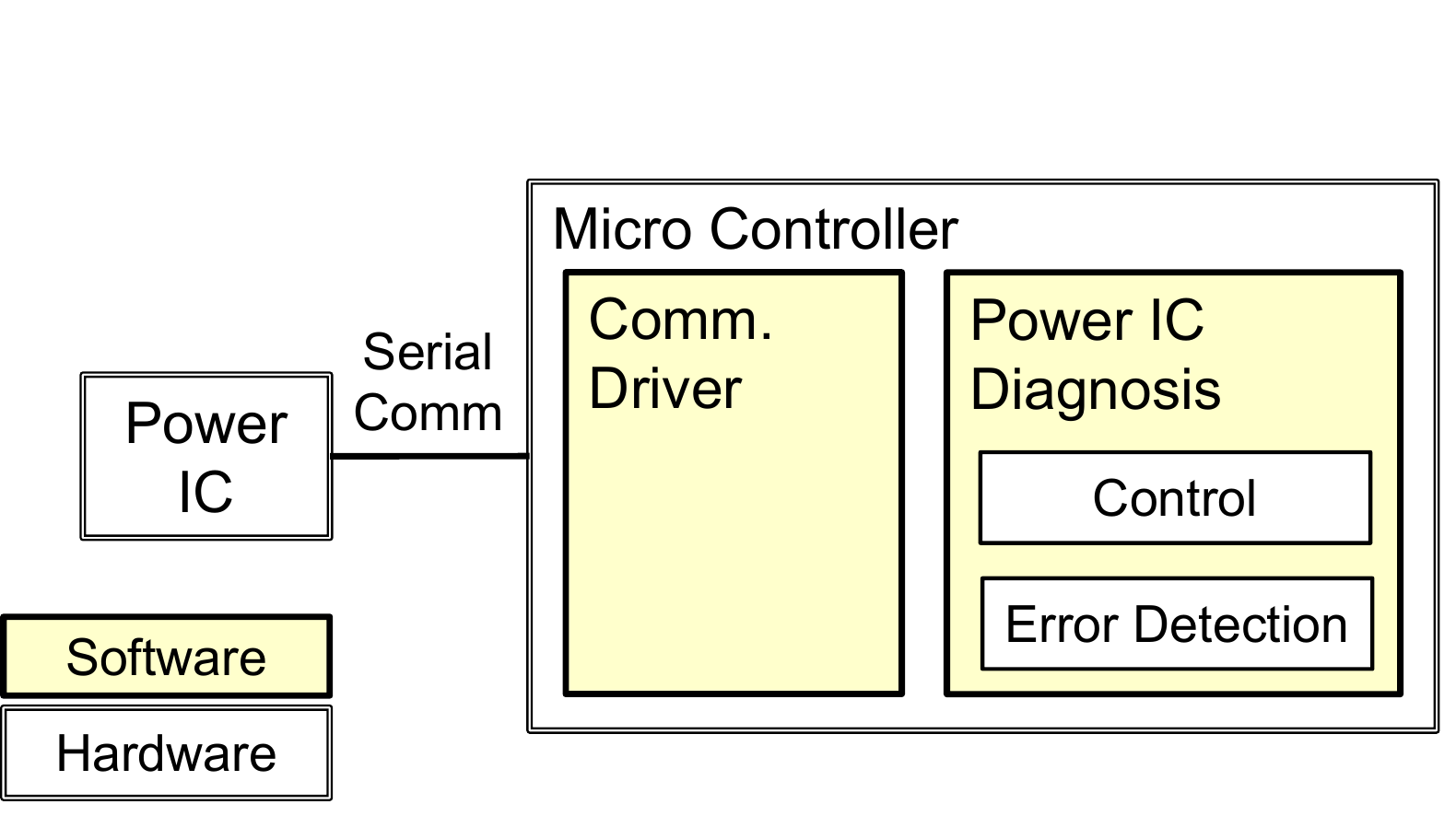}
  \caption{System Structure of Sample Case}\label{fig:sample_system}
 \end{center}
\end{figure}

The target to be verified is automotive software for a diagnosis of a power IC in a controller, as shown in Figure \ref{fig:sample_system}. The purpose of the verification was to investigate the reason for a malfunction. The malfunction was an incorrect error detection in the diagnosis on a long-term test. The diagnosis reports an error, but no fault was found in the power IC. Standard test methodology was not enough to reproduce the symptom and to find the cause of the malfunction, so model checking was applied.

The verification was done in two steps. In the first step, we focused on the serial communication driver between the micro controller and the power IC, but no defects were found. In the second step, we modeled the software for the diagnosis and the power IC. The communication driver was simplified as an external environment model, because it is proved the driver has no defect in the first step.

\begin{figure}[tb]
 \begin{center}
  \includegraphics[viewport=30 30 210 670, scale=0.6, angle=270, clip]{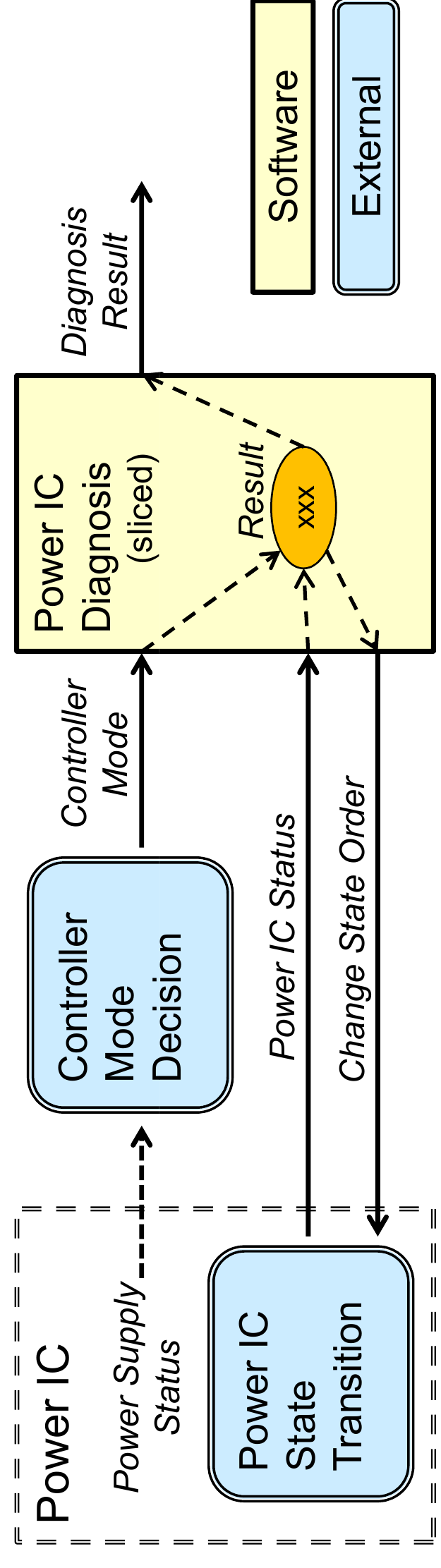}
  \caption{Model of Sample Case}\label{fig:sample_model}
 \end{center}
\end{figure}

The modeling method was applied in the second step. Figure \ref{fig:sample_model} shows the model. For the slicing, the variable which shows the result of the diagnosis was selected. The boundaries were adjusted so that the communication driver and some other part of the software were omitted. One of the inputs given by external environment models is the power IC status to be diagnosed, and another is the controller mode which is decided in the omitted software depending on the power supply. In the transition of the controller mode, a path was added that transits from the power-on state to the power-off state suddenly and abnormally. This path is intended to express the effect of an instantaneous drop of the voltage in the power supply.

In model checking, it is important to reduce the number of concurrent processes in order to reduce the state number. Although the micro controller and power IC run concurrently, the software model and the external model was set in the same one process, and the external environment models change its states before the diagnosis which is a cyclic task. If no malfunction was detected in this model, the external environment models could be separated into a different process. Furthermore, the data mapping was applied to the power IC status to be diagnosed. The sizes of the source code and models are shown in
Table \ref{tbl:sizes}; logical LOC refers to non-comment lines of code.

\begin{table}[htb]
 \begin{center}
  \begin{tabular}{ll}
   \hline \hline
   Item & Size (approx.) \\ \hline
   Whole source code (.c, .h) & 130 k Logical LOC \\ \hline
   Software model (Promela) & 600 LOC \\ \hline
   System model (Promela) & 750 LOC \\ \hline
  \end{tabular}
  \caption{Sizes of Source code and Model}\label{tbl:sizes}
 \end{center}
\end{table}

The result was that one defect was detected in the software which could cause a malfunction when the controller mode turns to the power-off state suddenly while the diagnosis software is in the specific state. It is confirmed that this malfunction could happen in the actual controller.

The mechanism of the malfunction is as follows. Although the diagnosis is stopped by power off, the power IC continues to change its state while the power is off. After the power supply is restored, an inconsistency between the actual power IC status and the recognition for that status by the diagnosis software occurs. This symptom happens only in a specific situation, so this is a timing problem. It is difficult to detect this defect with usual tests or simulations like HILS because a verifier can hardly set conditions to make this malfunction occur.

This case shows that model checking is effective against timing problems, and the modeling method can treat large-scale control software.
An obstacle to applying this modeling method to a development process is that it is time consuming. To improve the efficiency of the method, we developed a software tool to automate the method of generating verification models.

\section {Tool}
\label{sec:tool}

\subsection{Functionalities}
\begin{figure}[tb]
 \begin{center}
  \includegraphics[viewport=30 30 470 725, scale=0.65, angle=270, clip]{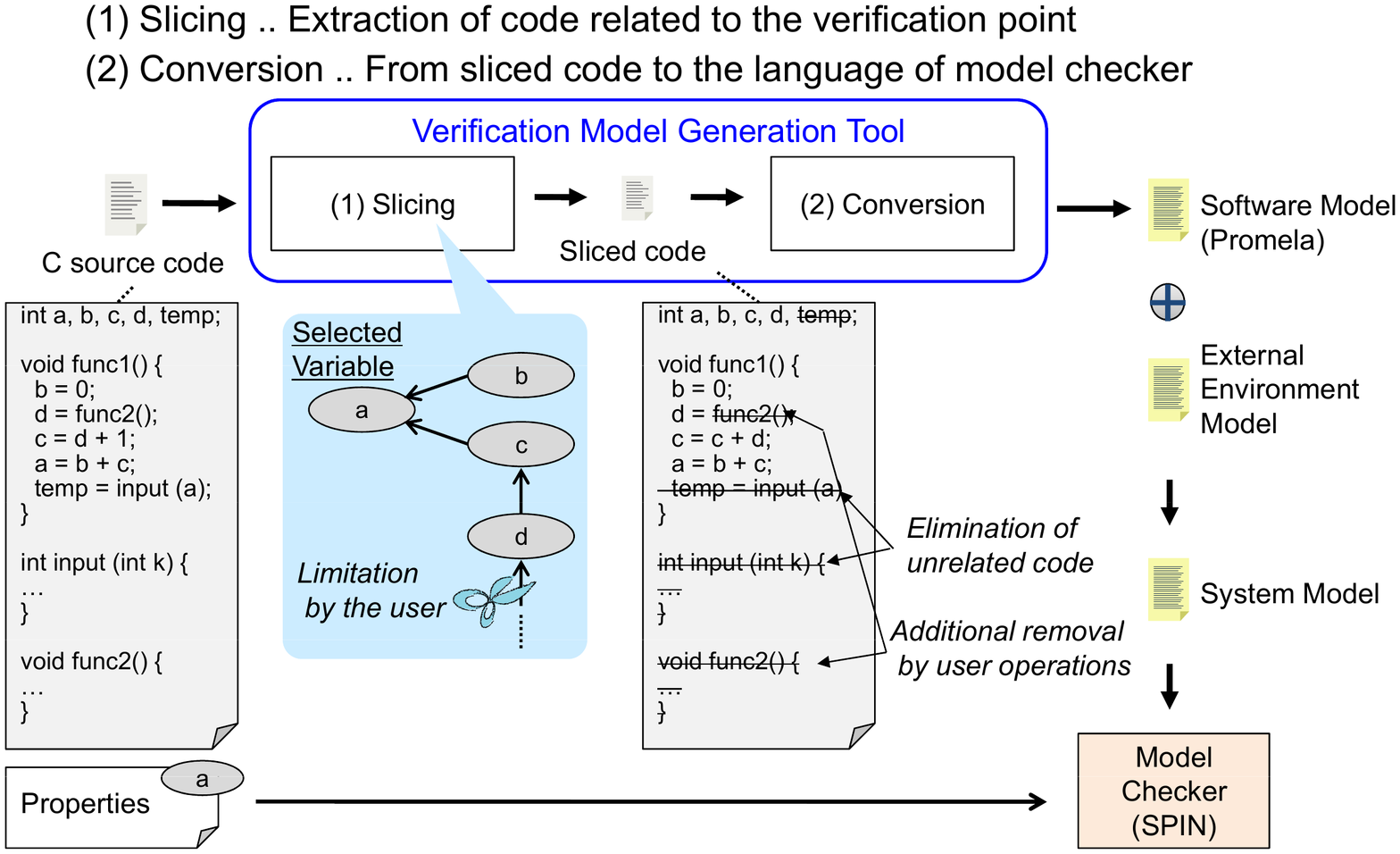}
  \caption{Verification Model Generation Tool}\label{fig:tool}
 \end{center}
\end{figure}

The tool we developed has two main functions as shown in Figure \ref{fig:tool}. One is slicing C language code based on a variable dependence graph; another is conversion from C language code to Promela.

When a user indicates one or more functions as execution start points of the program, the tool will analyze the variable dependencies. A variable dependence graph in a tree format will be extracted and displayed according to the variables that the user has selected. The tool can extract variable dependencies which extend to other execution paths, so that the effects of interruptions or other tasks can be considered.

\begin{figure}[!b]
 \begin{center}
  \includegraphics[viewport=25 30 450 600, scale=0.7, angle=270, clip]{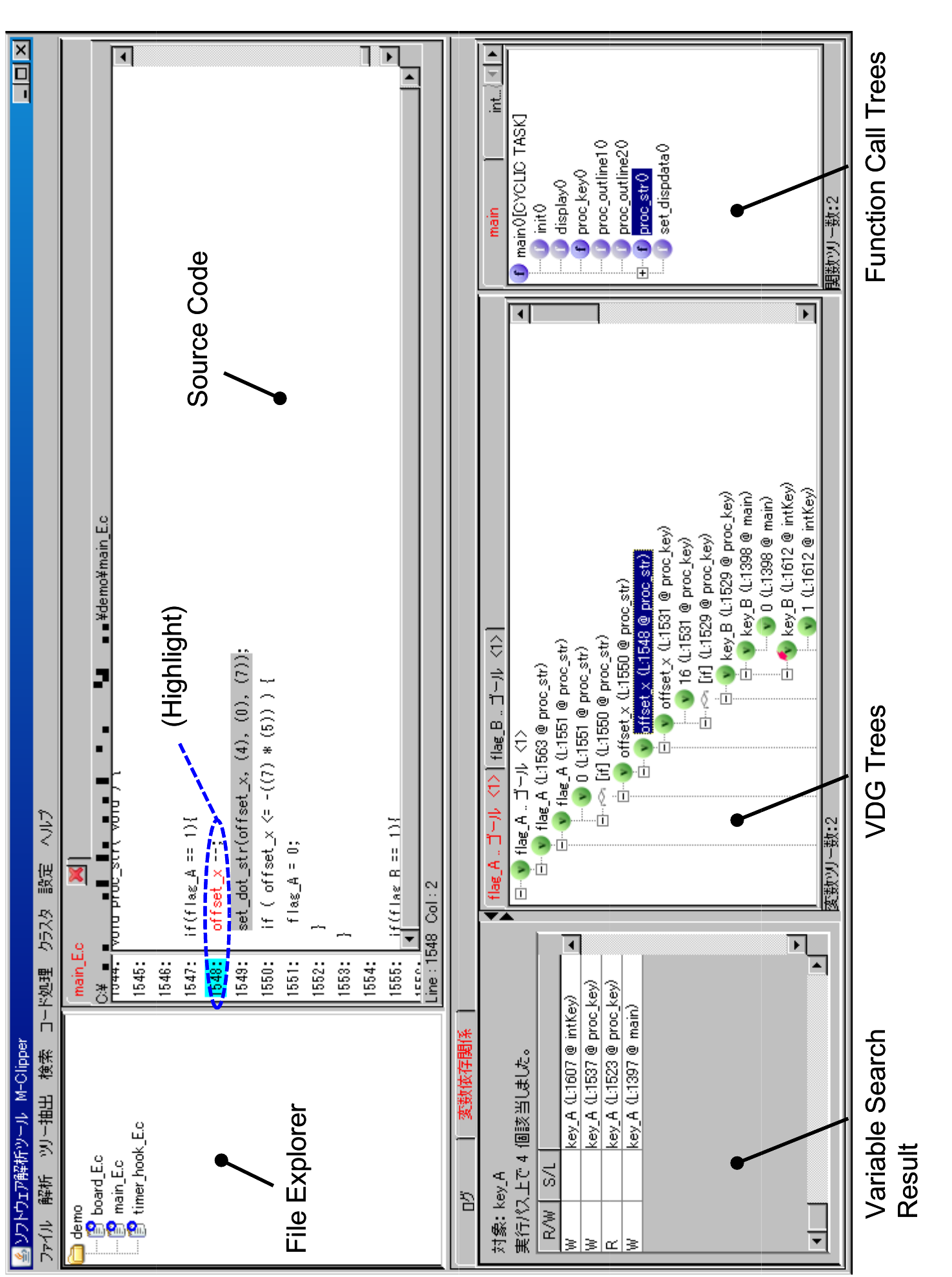}
  \caption{Screen Shot}\label{fig:screenshot}
 \end{center}
\end{figure}

The user can limit the variable dependencies to adjust the boundaries between the software models and external environment models. Such an operation is reflected as a color classification of the source code.
In the tool, variable dependence trees, function call trees, and source code files are displayed in association with each other, as shown in Figure \ref{fig:screenshot}. When a user selects a variable node on the VDG trees, related function node and code will be highlited to help the user understand the structure of the program, and to help the user to decide the boundaries.

In a tree format, the VDG could become larger than in a network format, so there are some refinements in the projection of the tree. Redundant tree paths (the same partial trees as other parts) are replaced with a node which indicates redundancy. Every node can be gathered to the function node on which the variable is written, so that a boundary adjustment between functions becomes easier. Moreover, the tool has search functions for variables on VDG trees or execution paths to help operations of the user.

At present, the limitations of the tool are as follows. Source code has to be pre-processed before the input. Pointers can be analyzed, but only one variable is pointed by a pointer, and pointed variables given by loop controls are ignored. Function calls by function pointers are also ignored. Slicing and model conversion are separated into different tools, and model conversion is done for each C code file. This tool is not open to the public.

\subsection{Algorithm for Analysis of VDG}
\begin{figure}[tb]
 \begin{center}
  \includegraphics[viewport=25 30 450 470, scale=0.6, angle=0, clip]{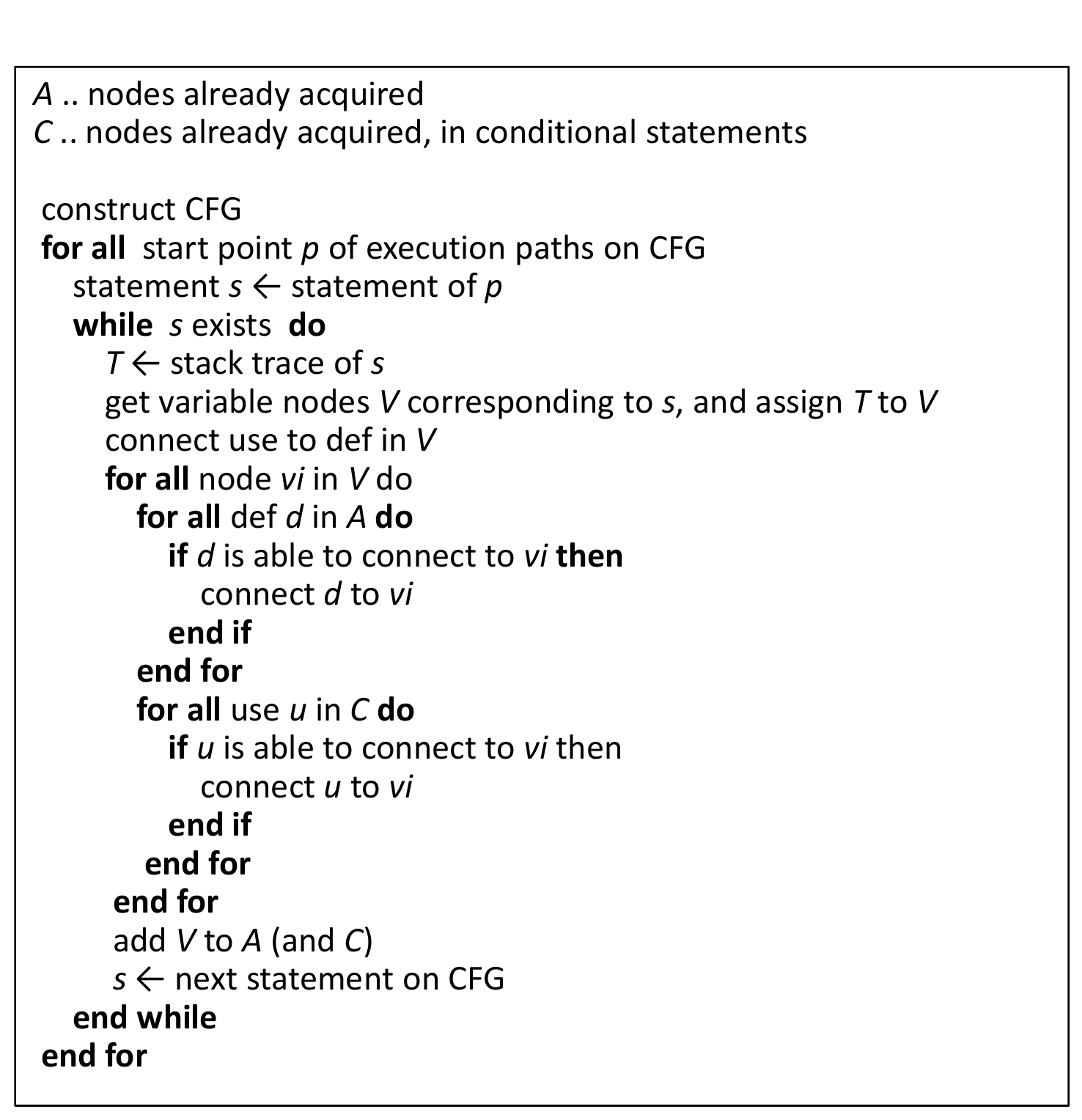}
  \caption{Algorithm for Analysis of VDG}\label{alg:alg1}
 \end{center}
\end{figure}

The algorithm to analyze VDG used in this tool is shown in Figure \ref{alg:alg1}.
Dependencies are connected one by one along the control flow graph.
Every node, which is a variable distinguished with the position in code and the execution path, is assigned information about the stack trace and associated to the control flow graph (CFG). Whether the nodes connect or not is judged using that information of the dependency source node, the sink node, and the nodes that have been already connected to the sink node.
Candidate nodes to connect obtained in {\it A} and {\it C} can be reduced in consideration of the attributes of the node to be connected, such as the substance (memory area) and the position in the CFG.
Dependencies due to loop statements are analyzed in the similar way.
Different from \cite{Horwitz:1988} \cite{Reps:1994}, initialization vertexes or finalization vertexes are not used.
The advantage of this algorithm we think is scalability. Actually, analysis was successful for code of which the size is over 350k logical LOC.

\begin{figure}[tb]
 \begin{center}
  \includegraphics[viewport=20 40 510 695, scale=0.6, angle=270, clip]{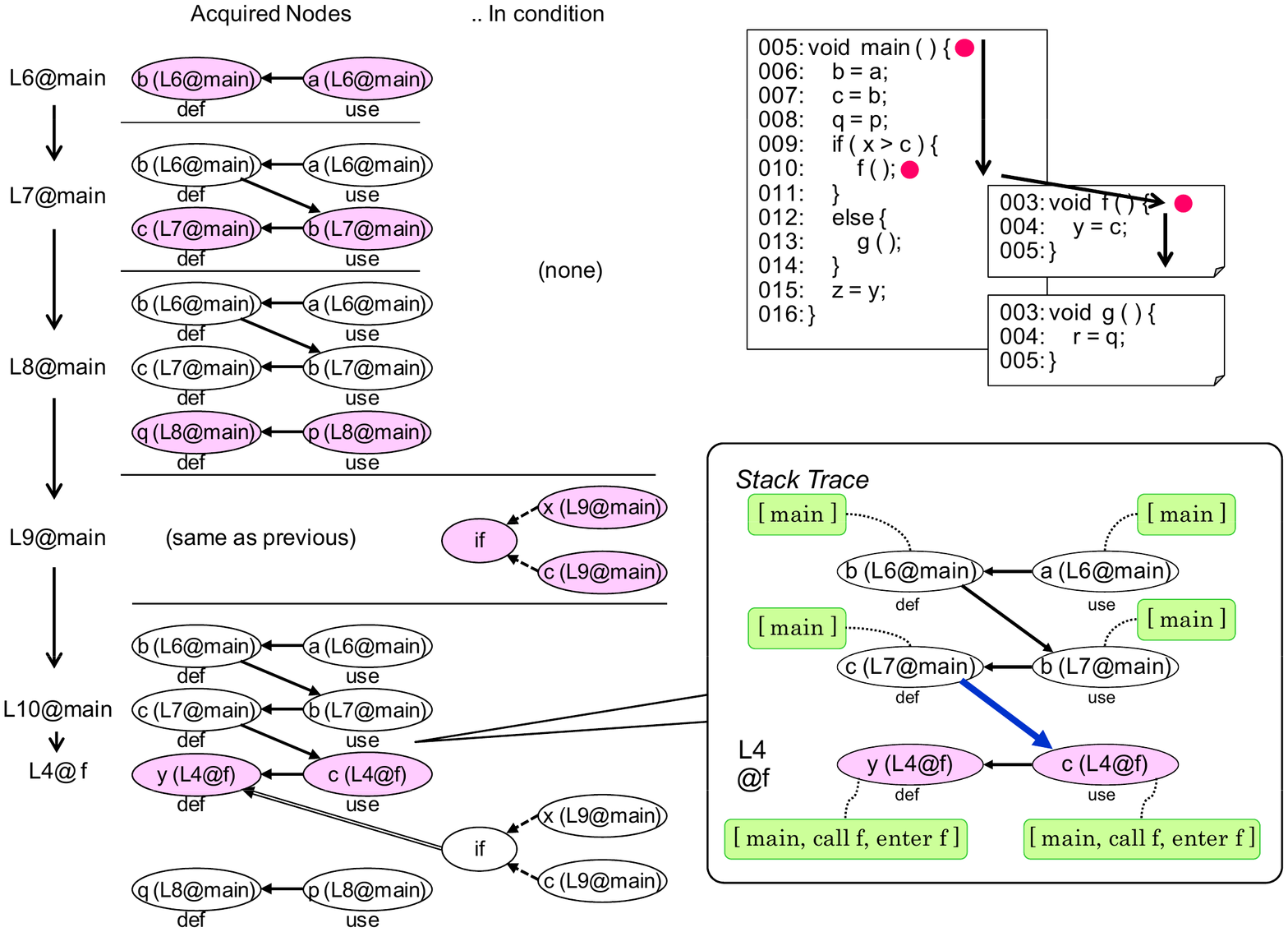}
  \caption{VDG Analysis Steps of Sample Code}\label{fig:analysis}
 \end{center}
\end{figure}

The process of the analysis for the sample code is shown in Figure \ref{fig:analysis}, where nodes in a conditional statement belong to a ``if'' node for the simplicity of appearance.
There are no function parameters and arguments in the sample code, but they can be treated in the same way.

After the analysis, a VDG for an execution path is obtained. This VDG becomes a network if there is a loop statement.
Goal/start trees are extracted from VDG with a selection of the variables, considering shared variables between different execution paths.

\subsection{Evaluation}
\begin{figure}[tb]
 \begin{center}
  \includegraphics[viewport=30 30 490 815, scale=0.5, angle=270, clip]{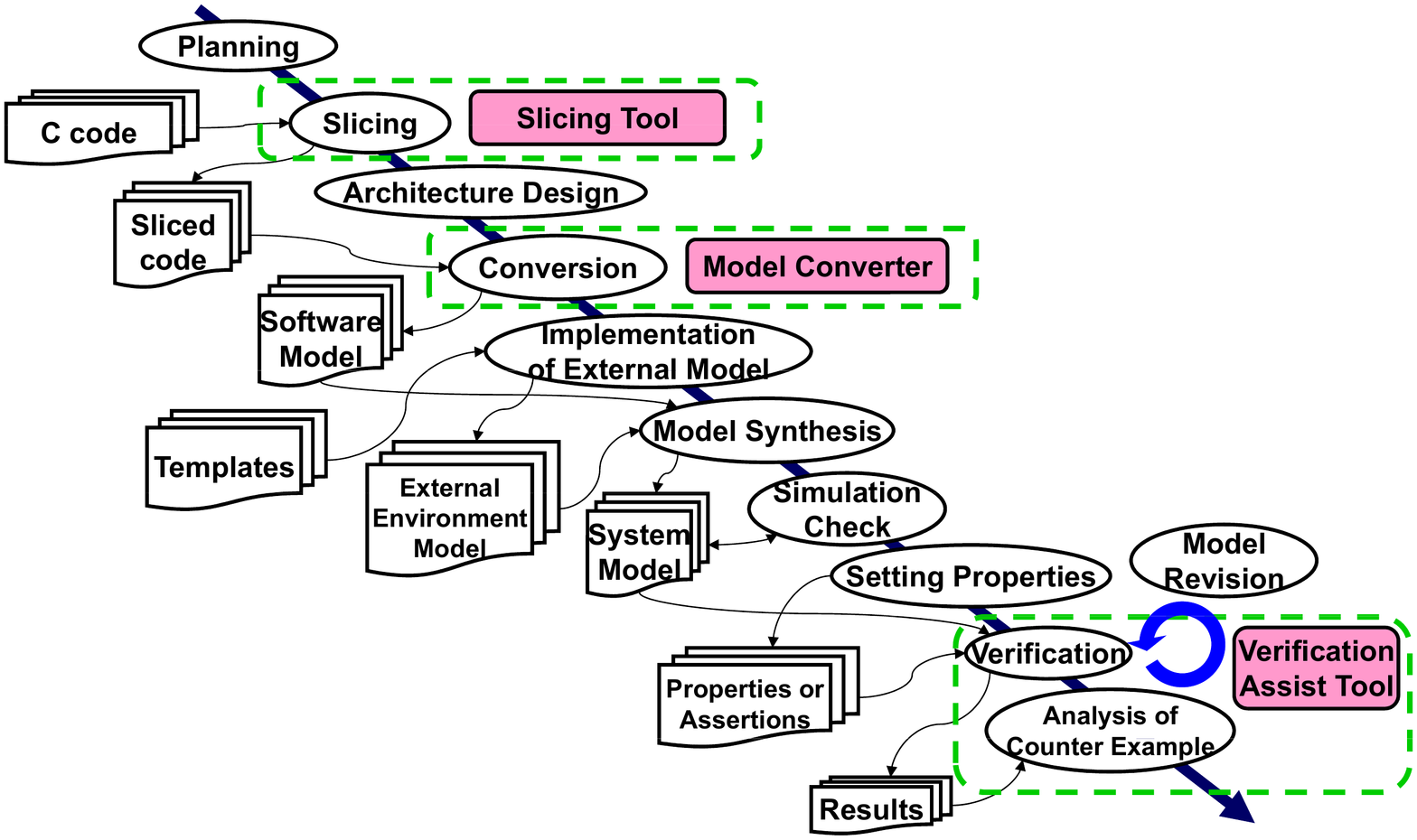}
  \caption{Verification Process}\label{fig:process}
 \end{center}
\end{figure}

Figure \ref{fig:process} shows the verification process with model checking we employed. Besides the slicing tool and the model converter, the verification assist tool was used in the process. This tool actuates checking on more than one property simultaneously, and shows a variable value table of a counter example in time series to help the user understand the mechanism that the violation of the property happens.
If a false positive / false negative error was detected in the verification, the step ``Model Revision'' is taken additionally to refine the model, and another verification will be done.

Table \ref{tbl:condition} shows the condition of the experiment. The target is automotive control software, and it had never been verified with model checking. Verifications were done without the tools first, and with the tools next.
Table \ref{tbl:time} shows measured time, and those values are normalized, as the total time in "without tools" is 1.
Highlighted rows are the steps to which the tools were applied.
The same measured time as the first try is set for some steps in the second try which should have no effects of the tools, because the verifier could work faster the second time.
The step ``Model Revision'' was taken because there was a false positive error at first.

\begin{table}[!htb]
 \begin{center}
  \begin{tabular}{ll}
   \hline \hline
   Item & Description \\ \hline
   Model checking experience of verifier & Approx. 1 year \\ \hline
   Target & Diagnosis of input signal \\ \hline
   Source code size (.c, .h) & 94 k Logical LOC \\ \hline
   Number of properties & 2 \\ \hline
  \end{tabular}
  \caption{Condition of Experiment}\label{tbl:condition}
 \end{center}
\end{table}

When tools are used, the whole verification time on model checking is reduced 35\%
compared to the verification without the tools. Although no tool is used in the step ``Simulation Check'' in which the verifier confirms the correctness of the model with simulation, the time is reduced because the model conversion by the tool is precise. There is no reduction of time in the step ``Verification'' because the number of properties is only two in this case. For the four steps in which the time reduction is obtained, it is expected that more reduction will be achieved when the tools are improved.

\begin{table}[!htb]
 \begin{center}
  \includegraphics[viewport=40 675 390 790, scale=1.2, angle=0, clip]{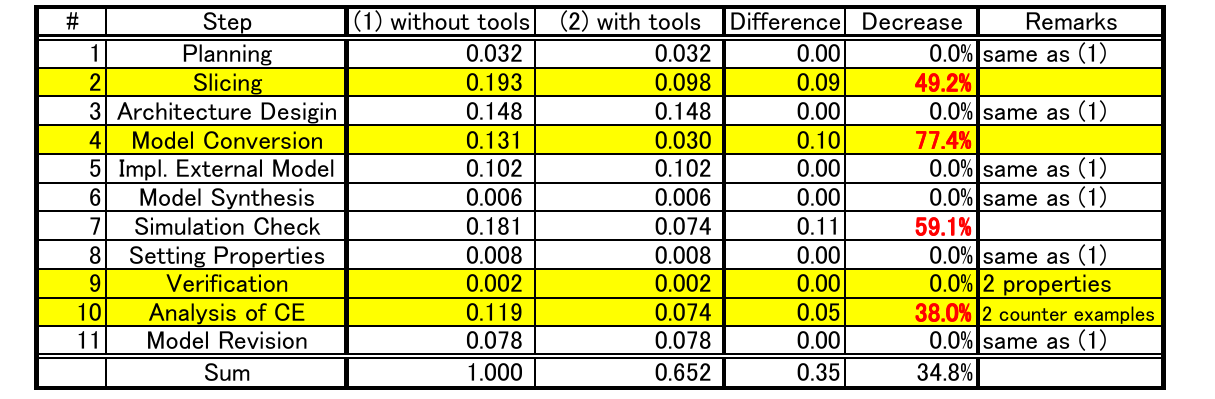}
  \caption{Result of Evaluation (execution times are normalized)}\label{tbl:time}
 \end{center}
\end{table}

Except for the user operations, the analysis and the slicing were completed within 5 minutes. (PC Spec: Intel(R) Core(TM)2 Duo E6550 @2.33GHz)


\section{Related Work}
\label{sec:related}
Bandera \cite{Corbett:2000} slices Java code and converts them into languages for model checker such as SPIN.
Modex / Feaver \cite{Holtzman:2005} slices C code using CFG, and converts it to Promela with a convert table and a test harness.
However, these tools provide no functionality to adjust an extent of the software to be verified.
The Wisconsin Program-Slicing Tool \cite{Reps:1994} \cite{Reps:1995} can analyze PDG / SDG, but it provides no functionality to adjust boundaries on PDG/SDG either.
CBMC \cite{clarke:2004}, BLAST \cite{Bayer:2007} and SLAM \cite{Ball:2011} input source code directly for model checking. SLAM verifies programs with software stubs as external environments. Such a static analyzer can detect defects automatically, but verification items are fixed.
It is important for a flexible verification to limit a target to be verified in the software, and the slicing technique in this paper enables it with less time than the ordinal slicing.

\section{Conclusion}
\label{sec:conclusion}
In embedded control systems, the potential risks of software defects have been increasing because of growing software complexity. To detect software defects which are difficult to be found with usual tests or simulations, we proposed a modeling method which can generate software models from source code for model checking, with a program slicing technique based on a variable dependence graph to avoid a state explosion.
VDG provides the measure to adjust a boundary between the software model converted from the source code and the external environment model, so that a large scale program can be verified with flexibility.
VDG also provides the information of interface variables between software models and external environment models, so that it becomes easier for a verifier to make or select external environment models to the software models.

We applied the proposed method to one case in automotive control software to find the cause of a non-reproducible malfunction, and showed the effectiveness of the method by clarifying that it was a timing problem caused by concurrent operations of the different hardware.
Furthermore, we developed the algorithm to analyze VDG, and we also developed a software tool to automate the generation of the model.
We achieved a 35\%
decrease in total verification time on model checking. It is expected that more time reduction will be achieved when the tool is improved.

\nocite{*}
\bibliographystyle{eptcs}
\bibliography{ftscs2012_Matsubara}
\end{document}